\documentclass[amsmath,amssymb,aps,prb,lengthcheck,floatfix]{revtex4}
\usepackage{graphicx}
\usepackage{dcolumn}
\usepackage{lmodern}
\usepackage{multirow}
\usepackage{hyperref}
\usepackage{subfigure}
\usepackage{color}
\usepackage{float}
\usepackage{pdfpages}

\begin{document}
	
	\title{Efficient Calculation of Excitonic Effects in Solids Including Approximated Quasiparticle Energies}
	
	\author{Filipe Matusalem}
	\email{filipematus@gmail.com, gmsn@ita.br}
	\author{Marcelo Marques}
	\author{Ivan Guilhon}
	\author{Lara K. Teles}
	\affiliation{Grupo de Materiais Semicondutores e Nanotecnologia (GMSN), Instituto Tecnol\'ogico de Aeron\'autica (ITA), 12228-900 S\~ao Jos\'e dos Campos/SP, Brasil}
	
	\begin{abstract}
		In this work we present a new procedure to compute optical spectra including excitonic effects and approximated quasiparticle corrections with reduced computational effort.  The excitonic effects on optical spectra are included by solving the Bethe-Salpeter equation, considering quasiparticle eigenenergies and respective wavefunctions obtained within DFT-1/2 method.
		The electron-hole ladder diagrams are approximated by the
		screened exchange.
		To prove the capability of the procedure, we compare the calculated imaginary part of the dielectric functions of Si, Ge, GaAs, GaP, GaSb, InAs, InP, and InSb with experimental data. 
		The energy position of the absorption peaks are correctly described. The good agreement with experimental results together with the very significant reduction of computational effort makes the procedure suitable on the investigation of optical spectra of more complex systems.     
		
	\end{abstract}
	
	\maketitle
	
	
	\section{Introduction}

	The density functional theory (DFT) is one of the most powerful and popular theoretical techniques to obtain groundstate and electronic properties of materials. However, it is well known that the excited spectra obtained from standard DFT suffer with the underestimation of energy transitions between Kohn-Sham (KS) eigenvalues. On the other hand, a good description of optical properties requires the knowledge of the excited states and a proper description of many-body effects. In this sense, solving the Bethe-Salpeter equation (BSE) for pair excitations and local-field contributions to the optical response can be taken into account\cite{Albrecht1998,Benedict1998,Rohlfing1998} to describe spectroscopic properties. Moreover, in some cases, such as the calculation of absorption spectra of low-dimensional materials, the inclusion of excitonic effects may dramatically affect the calculated results due to very large excitonic binding energy \cite{Wirtz2006,Guilhon2019}.
	
	In order to treat excited states, realistic quasiparticle (QP) energies are obtained by applying self-energy corrections to the KS energies, usually evaluated in the very demanding GW approximation\cite{Hedin1965}. The calculation of absorption spectra is often made in two steps. First, the QP energies and the RPA screened Coulomb interaction $W$ need to be obtained. In a second step the calculated eigenfunctions, together with the QP energies and $W$, are then used in the exciton calculation\cite{Albrecht1998}. The computation of the QP energies within the expensive GW method and the approximation of the electron-hole ladder diagrams by static screened Coulomb interaction, $W(\omega\to 0$), dramatically increase the time computational requirements and time. Therefore, large numerical effort is required to solve the BSE, restricting such calculations to the interaction of relatively few electron-hole pairs. Therefore, the study of low computational cost methods for obtaining an accurate optical spectrum considering excitonic effects is of great importance, and could open an avenue for studying more complex systems.
	
	In order to reduce the computational cost of the first step of the described procedure, in this work, we propose as an alternative the substitution of the GW QP energies by those obtained within DFT-1/2 method \cite{Ferreira2008,Ferreira2011}.  The DFT-1/2 method was developed to improve computational efficiency while keeping a good compromise with accuracy. This method is able to predict the energy gap results with precision similar to GW and HSE, but with the same computational effort of standard DFT calculations. It is worth mentioning, that this formalism is also free of adjusted parameters. In the sequence, the electron-hole ladder diagrams is then approximated by the screened exchange, provided by the hybrid HSE06\cite{Heyd2003,Heyd2004,Heyd2006} scheme. 
	
	
	To prove the accuracy of this methodology, we compute the imaginary part of dielectric function of a set of semiconductors, namely, Si, Ge, GaAs, GaP, GaSb, InAs, InP and InSb, and compare the obtained results with available experimental data. The article is organized as follows. In Sect. II, we describe the methods adopted and computational details of our calculations. In Sect. III, we present and discuss the results. Finally, in Sect. IV we briefly summarize the paper and present our conclusions.

	\section{Computational details}
	
	We perform DFT calculations employing a semi-local approach including a second-order gradient coefficient for the exchange energy named PBEsol\cite{Perdew2008,Perdew2009}. The PBEsol is employed to determine the theoretical lattice parameters and relaxed positions of ions. The Kohn-Sham equations are solved self-consistently within the projector augmented wave (PAW) for generation of wavefunctions and pseudopotentials \cite{Kresse1999}, as implemented in VASP. For the plane wave expansion of the electronic wave functions we use an energy cutoff of 245 eV, 174 eV, 209 eV, 255 eV, 283 eV, 209 eV, 255 eV and 172 eV for Si, Ge, GaAs, GaP, GaSb, InAs, InP and InSb, respectively. Spin-orbit coupling was took into account for all structures including Sb atoms, in calculation of band structures. The Brillouin zone (BZ) integrations are carried out with a 15$\times$15$\times$15 $\Gamma$-centered $k$-point mesh. 
	
	The general rule for the inclusion of the self-energy correction in the DFT-1/2 approach is to distribute the half-ionization between the atomic  orbitals forming the valence band maximum (VBM) according to the percentage of each orbital contribution \cite{Ferreira2008}. For GaAs, GaSb and InSb around 90\% of the VBM is formed by the $p$ orbital of the correspondent anion atom in each case. Therefore an accurate result is achieve with all the half-ionization included for the corresponding orbital. In Si and Ge, a $p$ character of VBM states and covalent bonds between atoms of same species are observed. In these cases the half-ionization is shared for each pair of atoms, so the the atomic potential is generated with only 1/4 ionization to avoid the overestimation of the self-energy correction on the QP eigenenergies. As described in the papers introducing the DFT-1/2\cite{Ferreira2008,Ferreira2011}, the true band gap is a result of a sum of the standard DFT Kohn-Sham band gap plus the self-energy contribution relative to the VBM, S$_V$, minus the self-energy contribution relative to the conduction band minimum (CBM), S$_C$. 
	
	For three-dimensional compounds in general, the contribution relative to the CBM can be ignored and a good description of the band gap already achieved by considering only corrections on the valence states. However, in the case of cubic germanium compound and GaP, the self-energy contribution relative to the CBM could not be ignored, mainly due the known existence of a indirect band gap. The self-energy S$_C$ correction is then made in this case applying the DFT-1/2 procedure for the Ge($s$), Ga($s$) and P($s$) orbitals forming the CBM. To adjust the sign of the contribution, in order to consider the addition of half electron to the unoccupied band, a negative amplitude is used\cite{Matusalem2015}. For InAs and InP, the half electron ionization is shared between both species according to the VBM composition. This is controlled by the amplitude parameter included in the generation of the DFT-1/2 PAW potential\cite{Ataide2017}. The details for the generation of the PAW corrected potentials used in this work are listed in Table \ref{tab2}. 
	
	Then, the  DFT-1/2 QP eigenenergies and respective wavefunctions are used to build a two-particle Hamiltonian of singlet excitations to describe excitonic effects. For given sets of valence bands, conduction bands, and k-points we solve the eigenvalue problem of the two-particle Hamiltonian, the homogeneous Bethe-Salpeter equation \cite{Bechstedt2015}. The dielectric functions are calculated considering both the independent quasiparticle and excitonic effects for comparison.

	\section{Results and Discussion}
	Lattice constants calculated using PBEsol exchange-correlation functional are listed in Table \ref{tab1}.  A very good agreement between the calculated results and experimental findings is observed, only exhibiting overestimations not much larger than 1\%. The resulting relaxed structures were used for the calculation of electronic and optical properties.       
	
	\begin{table}[htb]
		\centering
		\caption{\label{tab1} Theoretical PBEsol ($a_{PBEsol}$) and experimental\cite{levinshtein1996handbook} ($a_{EXP}$) lattice parameters for a set of semiconductors under study in this work. } 
		\begin{ruledtabular}
			\begin{tabular}{cll}
				& \bf $a_{PBEsol}$ (\AA) & \bf $a_{EXP}$ (\AA)   \\\hline
				Si     &    5.439       		& 		5.431	    	\\
				Ge     &    5.704       		&		5.658			\\
				GaAs   &    5.685       		& 		5.653			\\ 
				GaP    &    5.477       		& 		5.450			\\ 
				GaSb   &    6.115       		& 		6.096			\\
				InAs   &    6.123       		& 		6.058			\\  
				InP    &    5.932       		& 		5.869			\\ 
				InSb   &    6.543       		& 		6.479			\\
			\end{tabular}
		\end{ruledtabular}
	\end{table}
	
	Table \ref{tab2} shows the electronic band gaps calculated considering PBEsol functional and DFT-1/2 method, together with experimental data. As expected, the PBEsol band gaps are underestimated in comparison with experiment, including the wrong prediction of metallic behavior for Ge, GaSb, InAs, and InSb. After the inclusion of the self-energy correction following the DFT-1/2 method, no prediction of metallic behavior is observed and more accurate band gaps are obtained for the semiconductor materials. Good agreement between calculated and experimental results is achieved. 
	
	\begin{table*}[htb]
		\centering
		\caption{\label{tab2} PBEsol (E$_{PBEsol}$), DFT-1/2 (E$_{DFT-1/2}$) and experimental\cite{levinshtein1996handbook} (E$_{EXP}$) electronic band gaps. The last column show detailed information about the DFT-1/2 correction. The plus(minus) sign means correction applied to the valence(conduction) level. 0.5 and 0.25 are the amount of electronic ionization in the atomic calculation applied to $p$ or $s$ orbitals. The superscript $d$ and $i$ stands for direct and indirect gap, respectively.   } 
		\begin{ruledtabular}
			\footnotesize
			\begin{tabular}{cclcc}
				& \bf E$_{PBEsol}$ (eV) & \bf E$_{DFT-1/2}$  (eV) & \bf E$_{EXP}$ (eV) &   \bf Atom(Orbital) fraction CUT Amplitude   \\\hline
				Si$^i$ &         0.47          & 1.23                 &        1.12        &             Si(p) 1/4 3.78 1.00              \\
				Si$^d$ &         2.52          & 2.94                 &        3.40         &             Si(p) 1/4 3.78 1.00              \\
				Ge$^i$ &         metal         & 0.77                 &        0.66        & Ge(p) 1/4 4.23 1.00 and Ge(s) 1/4 2.27 -1.00 \\
				Ge$^d$ &         metal         & 0.80                 &        0.80         & Ge(p) 1/4 4.23 1.00 and Ge(s) 1/4 2.27 -1.00 \\
				GaAs  &         0.39          & 1.38                 &       1.42        &             As(p) 1/2 3.84 1.00              \\
				GaP$^i$ &         1.51          & 2.39                 &        2.26        &  P(p) 1/2 3.70 0.82, P(s) 1/2 1.90 -0.5 and Ga(p) 1/2 1.10 0.18, Ga(s) 1/2 2.70 -0.5  \\
				GaP$^d$ &         1.69          & 2.97                 &        2.78        &  P(p) 1/2 3.70 0.82, P(s) 1/2 1.90 -0.5 and Ga(p) 1/2 1.10 0.18, Ga(s) 1/2 2.70 -0.5  \\
				GaSb  &         metal         & 0.66                 &       0.73        &             Sb(p) 1/2 4.27 1.00              \\
				InAs  &         metal         & 0.51                 &       0.35         & As(p) 1/2 4.14 0.86 and In(p) 1/2 1.36 0.14  \\
				InP   &         0.48          & 1.42                 &       1.34         & P(p)  1/2 4.00 0.86 and In(p) 1/2 1.36 0.14  \\
				InSb  &         metal         & 0.36                 &        0.17        &             Sb(p) 1/2 4.53 1.00              		\end{tabular}
		\end{ruledtabular}
	\end{table*}
	
\begin{figure*}[ht]
		\includegraphics[width=0.88\linewidth]{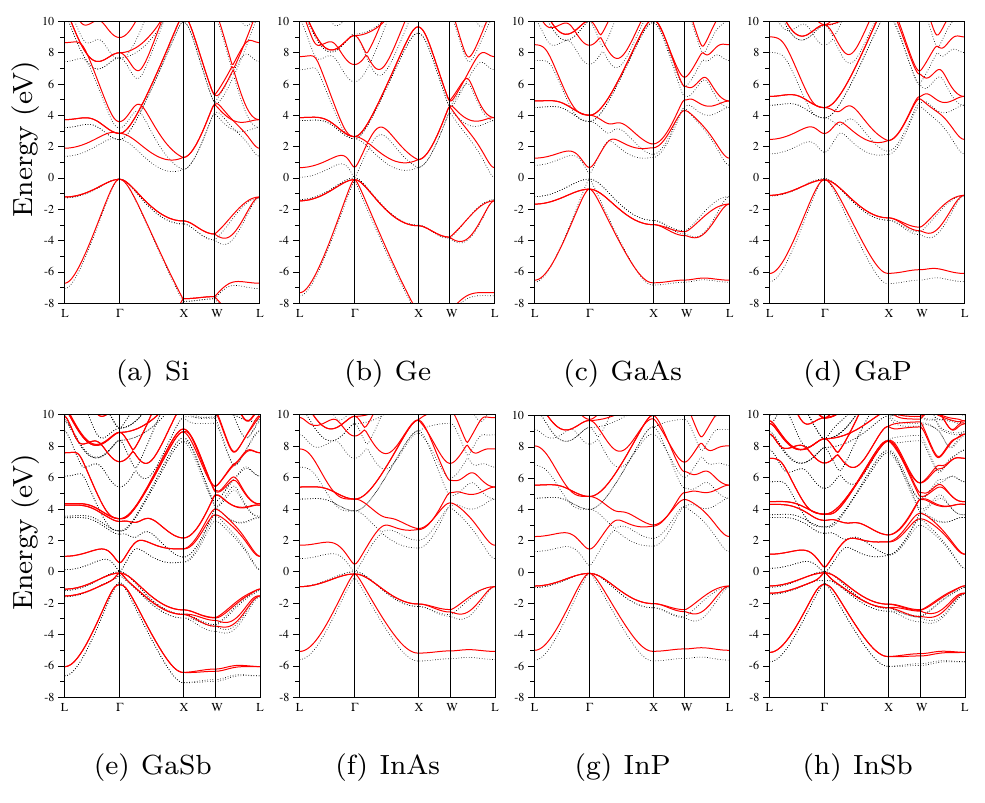}
		\caption{\label{fig1}(Color online) PBEsol (black dotted lines) and DFT-1/2 (red straight lines) band structures. The Fermi level is chosen as zero energy. For GaSb and InSb spin-orbit coupling is included.}
	\end{figure*}

	In Fig. \ref{fig1} are show the band structures calculated using PBEsol and DFT-1/2 approaches. We clear see the opening of the band gaps due the inclusion of the DFT-1/2 correction, corroborating the data of Table \ref{tab1}. Also, a known characteristic of DFT-1/2 can be observed, the trend to flattening the bands. The effect occurs mainly on the valence bands since are the ones formed by the atomic orbitals subject to the applied half-ionization (see last column of Table \ref{tab2}).    
	
\begin{figure*}[hptb]
		\subfigure[Si]{\includegraphics[width=0.3\linewidth]{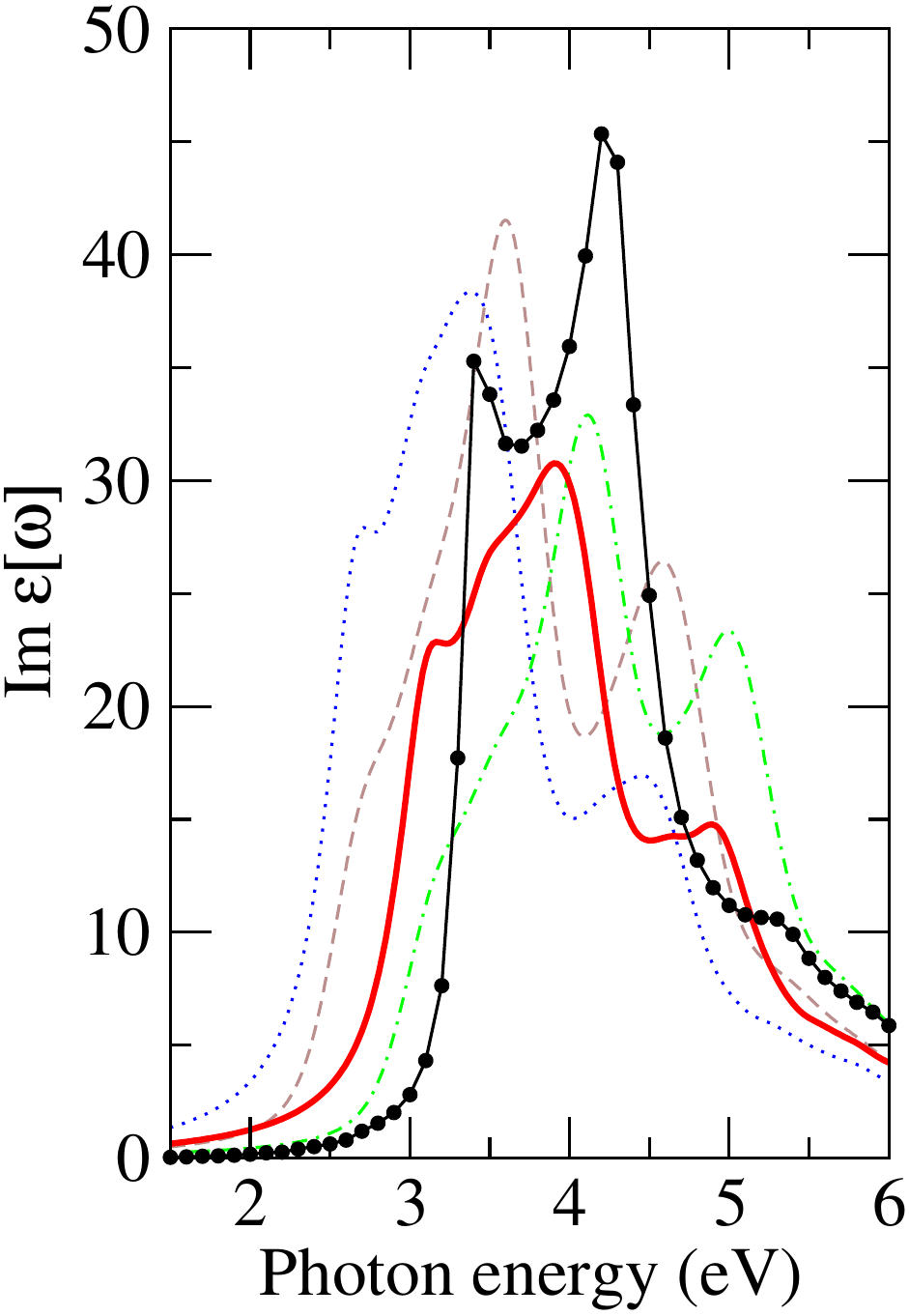}}
		\subfigure[Ge]{\includegraphics[width=0.3\linewidth]{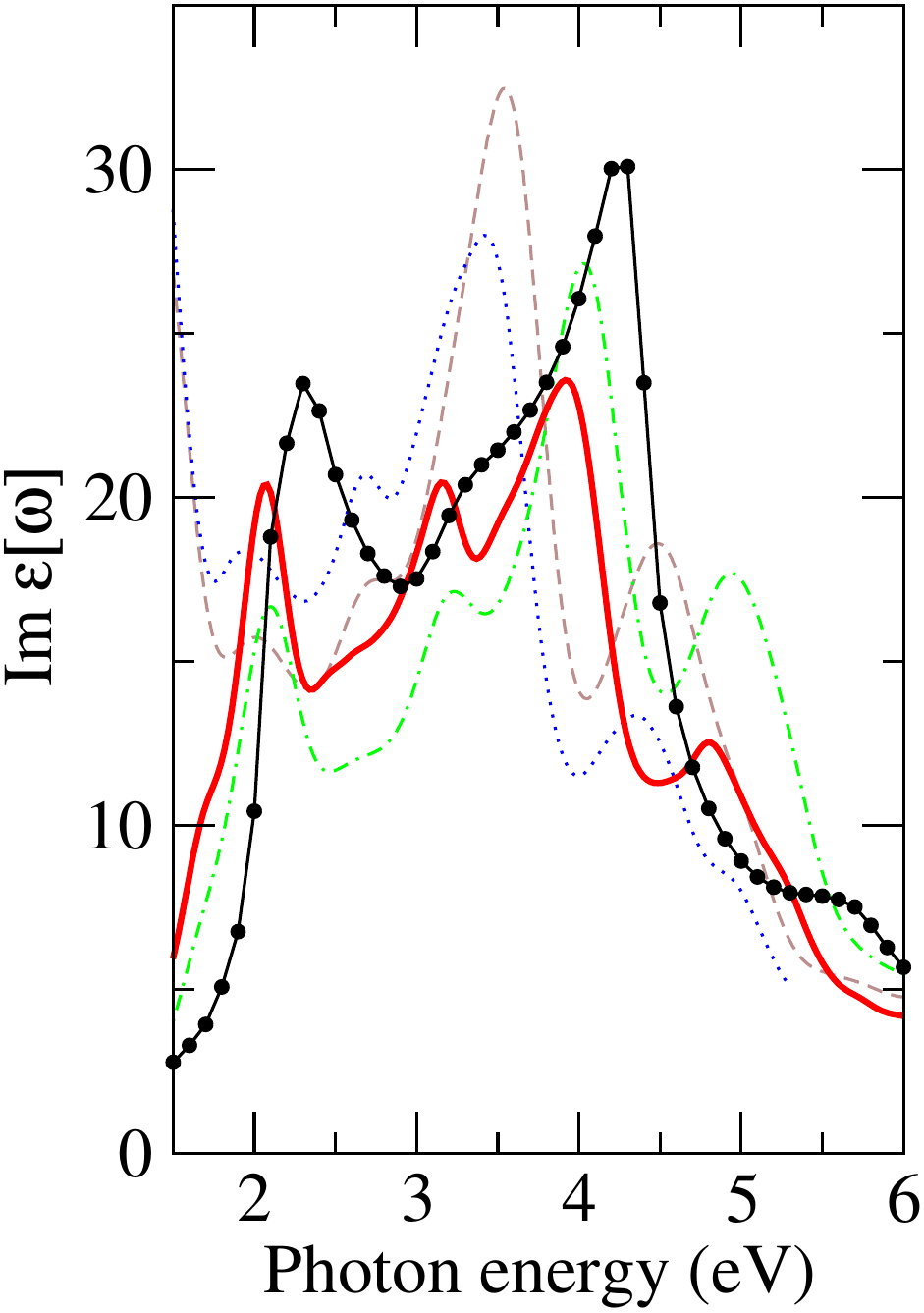}}
		\subfigure[GaAs]{\includegraphics[width=0.3\linewidth]{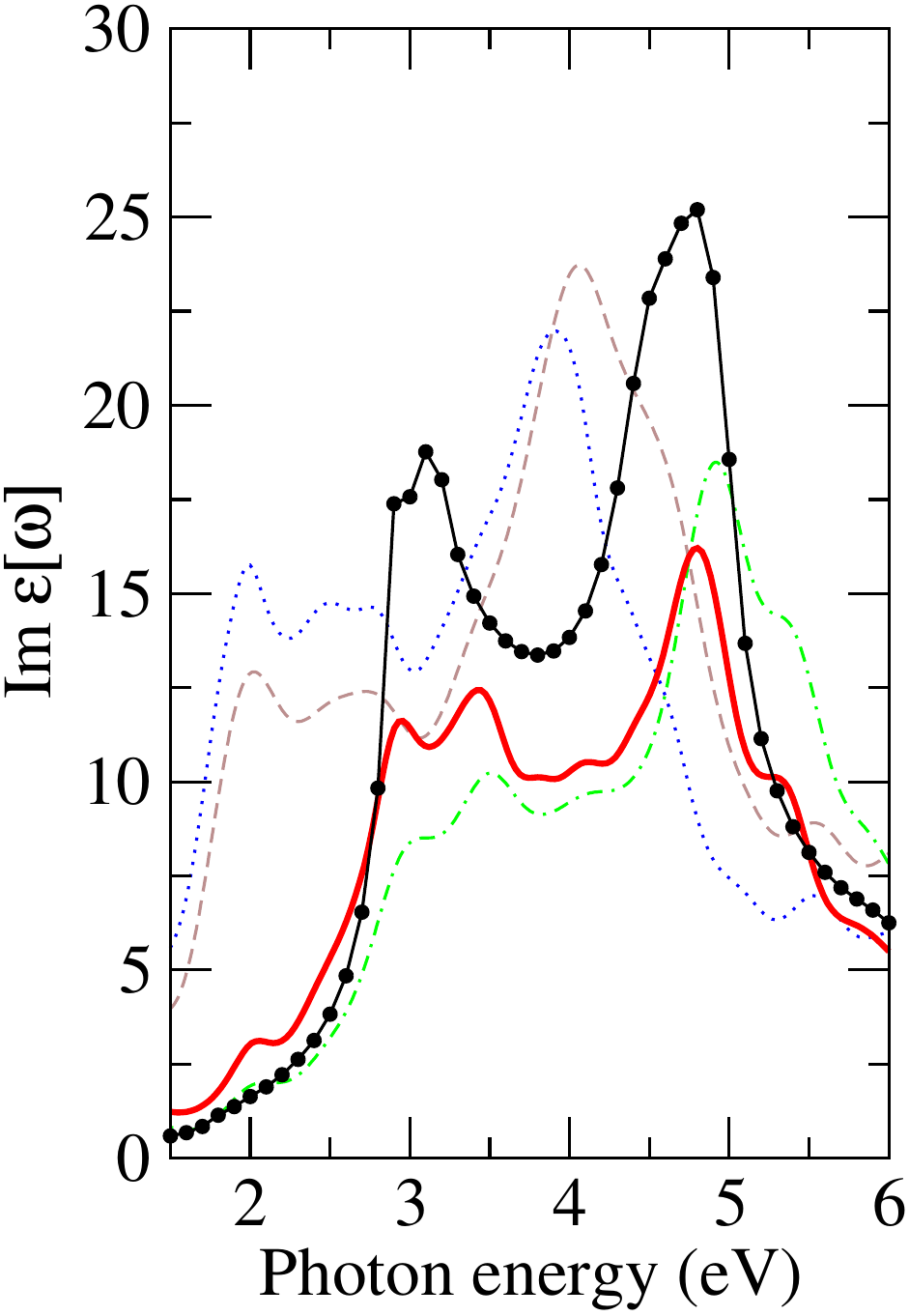}}
		\subfigure[GaP]{\includegraphics[width=0.3\linewidth]{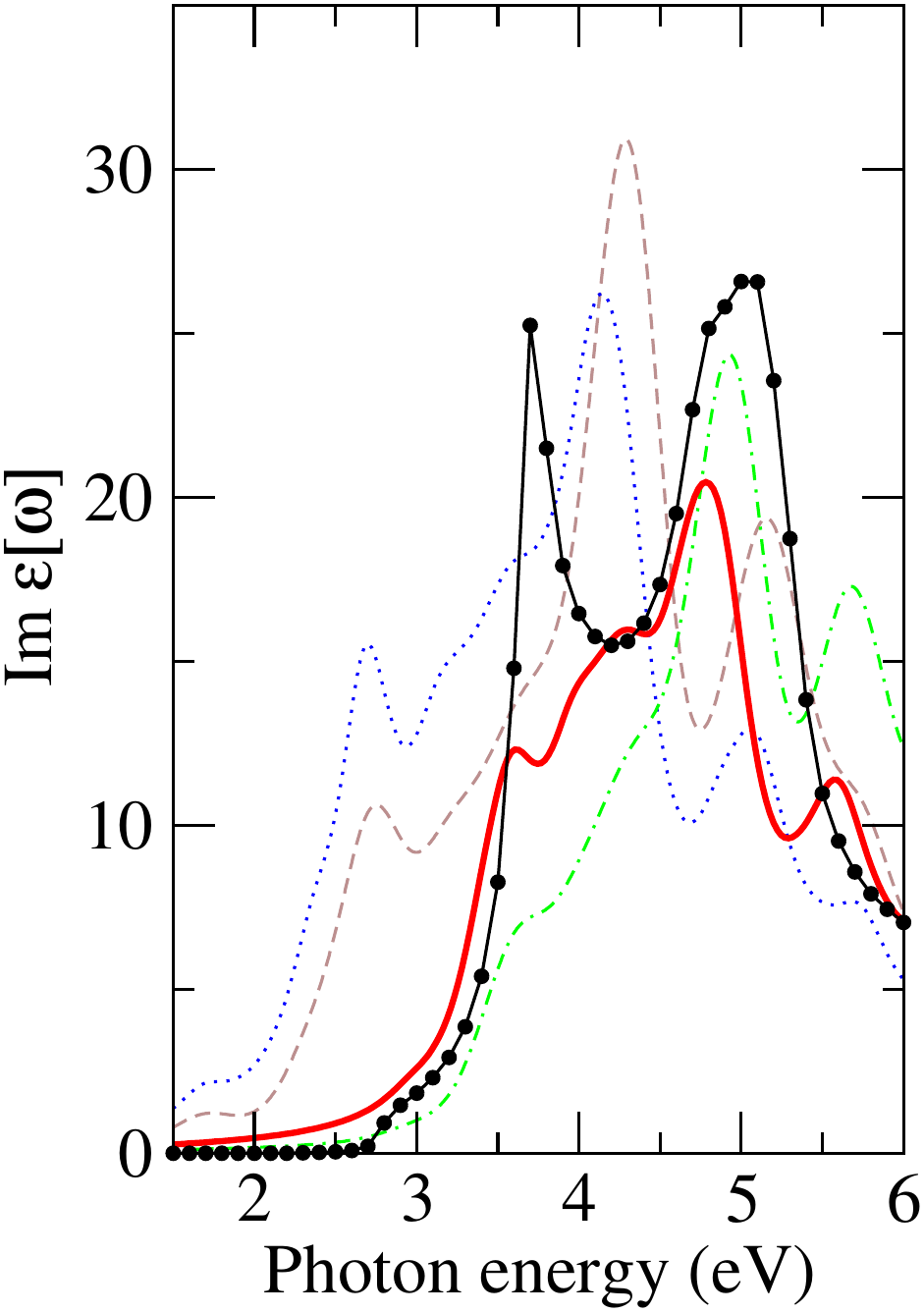}}
		\subfigure[GaSb]{\includegraphics[width=0.3\linewidth]{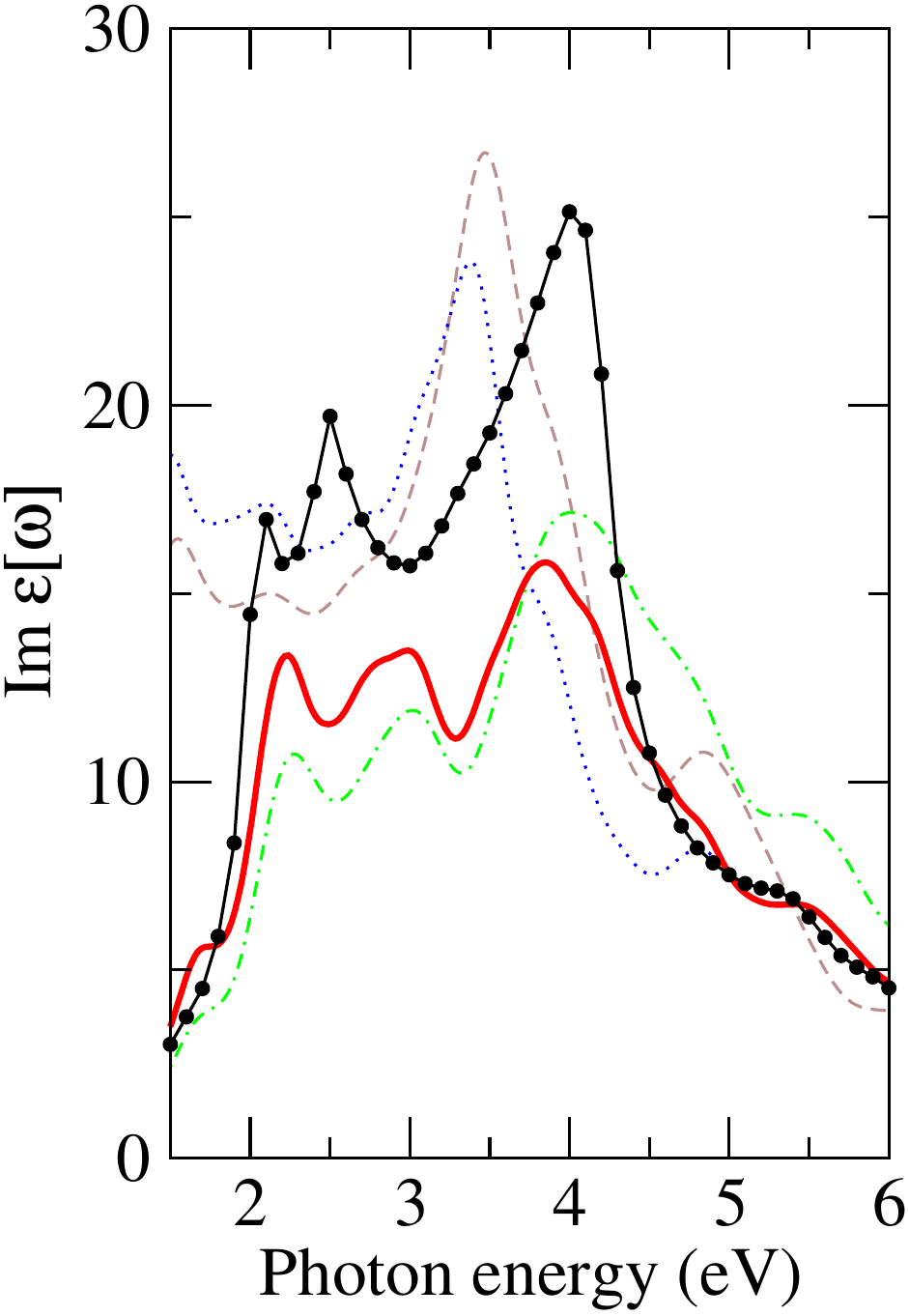}}
		\subfigure[InAs]{\includegraphics[width=0.3\linewidth]{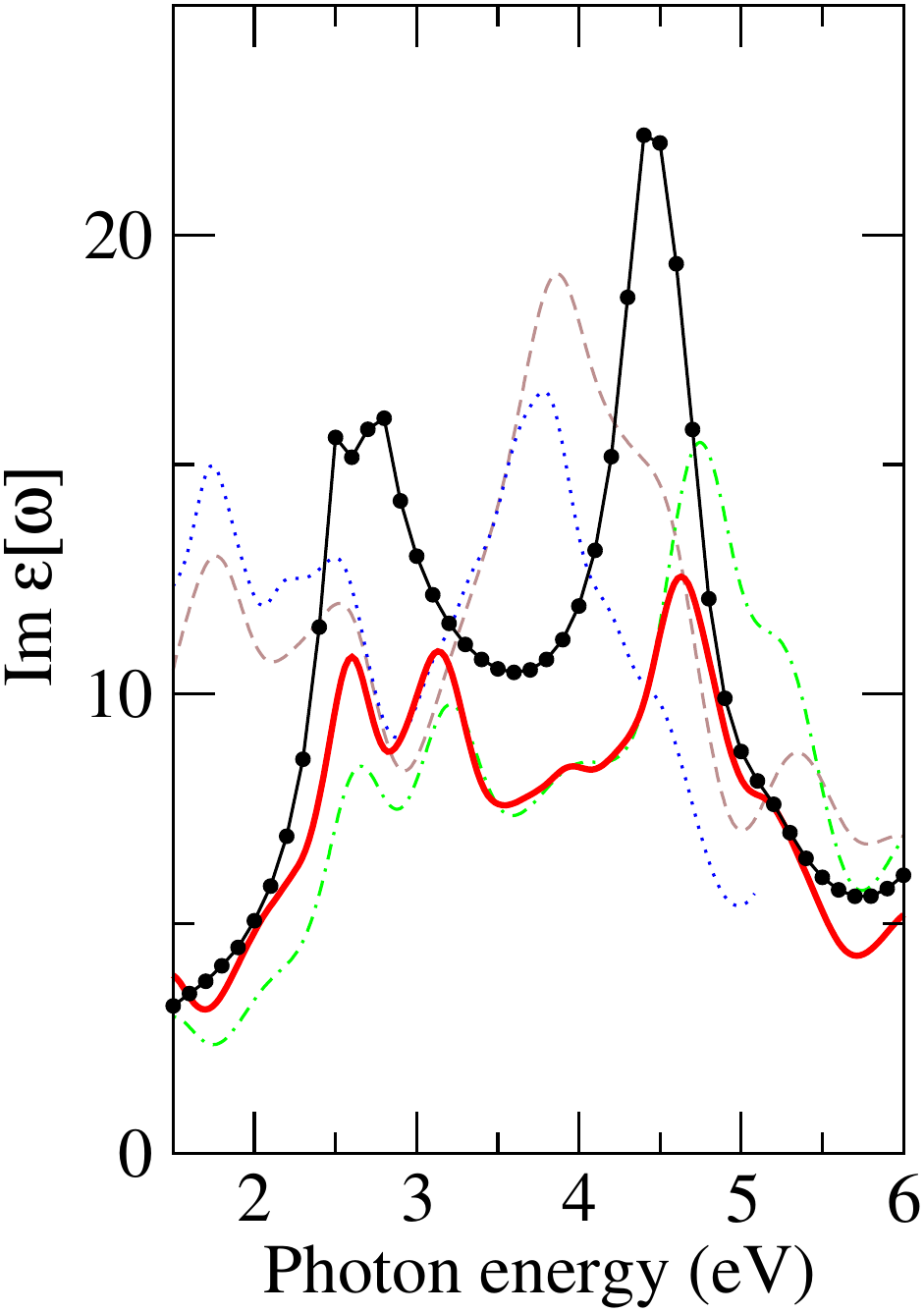}}
		\subfigure[InP]{\includegraphics[width=0.3\linewidth]{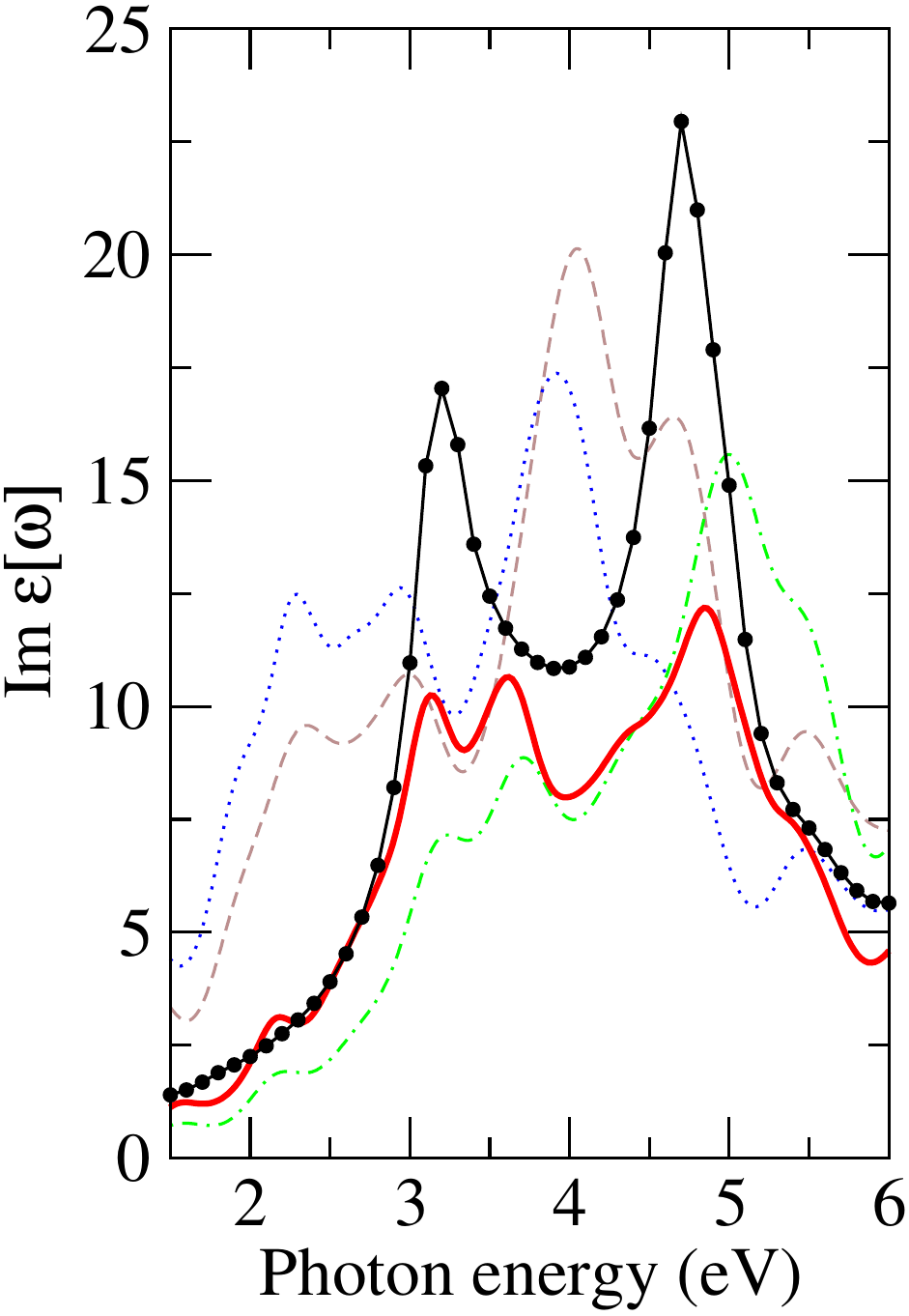}}
		\subfigure[InSb]{\includegraphics[width=0.62\linewidth]{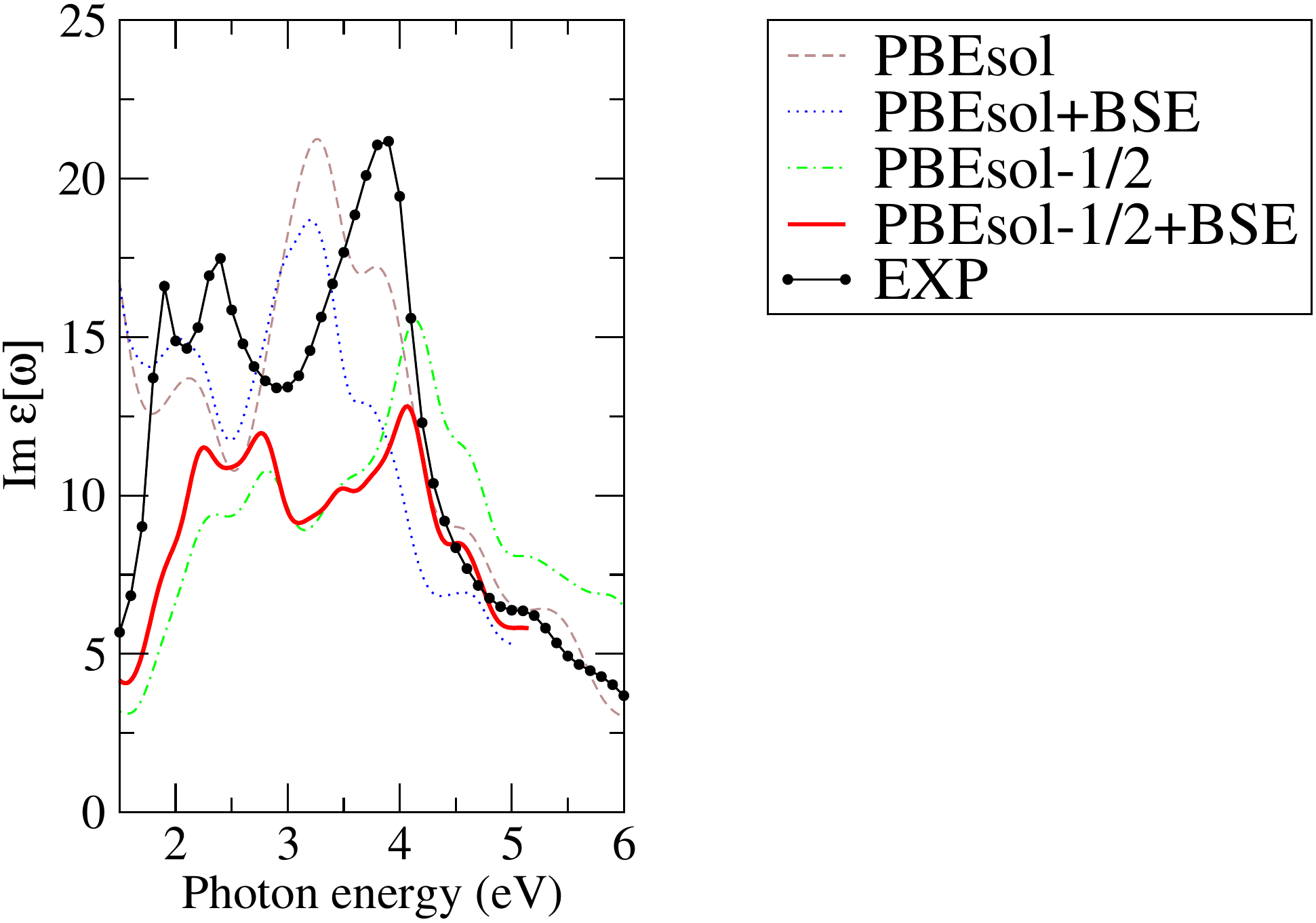}}
		\caption{\label{fig2}(Color online) Imaginary part of dielectric function. Gray dashed curve: pbesol. Blue dotted curve: PBEsol+BSE. Green dot-dash curve: PBEsol-1/2. Red continuous curve: PBEsol-1/2 + BSE. Black line-Dots curve: experiment\cite{Aspnes1983}. }
	\end{figure*}

	The imaginary part of the dielectric function, calculated using PBEsol, PBEsol $+$ BSE, DFT-1/2, and DFT-1/2 $+$ BSE, are shown in Fig. \ref{fig2} in comparison with experimental results. At least four valence and four conducting bands are included in the calculation of optical spectra, which guarantee that transitions up to 10 eV are included, as can be inferred from band structures in Fig. \ref{fig1}. The obtained optical spectra are all depicted in \ref{fig2}. A common characteristic observed in all spectra is that the curves obtained without inclusion of QP corrections (PBEsol) are severe redshifted with respect to the experimental results. This behavior is justified by the underestimation of the fundamental band gap, since the imaginary part of the dielectric function is strictly related with the interband transitions from the valence to conduction bands. Considering the green curves in Fig.~\ref{fig1}, one observes that the inclusion of the DFT-1/2 QP correction shifts the previous PBEsol results to greater energies in the same rate that the ones observed for the increasing in the band gaps shown in Table \ref{tab2}. As for the band gaps, the inclusion of the QP correction resulted in imaginary dielectric functions in very good agreement with the experimental results when looking only for the position in energy of the main peaks present in the spectra. The exception, although improvement is achieved, occurs for the silicon case. As can be observed in Table \ref{tab2} the direct gap of silicon is still 0.4 eV underestimated in comparison with the experiment. This difference is approximately the same observed in the absorption spectra of silicon (Fig. \ref{fig1} (a)). The inclusion of the excitonic effects on the other hand yields in a small blueshift of the spectra. This is expected and can be associated with the decrease in band gap due the influences of the bound energy of the electron-hole pair (exciton) generated by the excitation of an electron from the valence to conduction band. Peaks in spectra related to bound excitons are not visible but could appear below the onset of the absorption if a much more refined k-point meshes is used\cite{Rohlfing1998,Hahn2005a,Hahn2005}. Another consequence of the inclusion of the excitonic effects is related with the intensities. Peaks located in the lowest energy of the spectra (to the left side) have their intensities decreased while peaks located in the highest energy of spectra (to the right side) have their intensities increased. The spectra of GaAs, GaSb, InAs, InP and InSb, presents a double peak centered at about 3.1, 2.5, 2.8, 3.4,  and 2.5 eV, respectively. The inclusion of the excitonic effects increase the intensity of both peaks, however, the left one that before had lower intensity becomes more pronounced than the right one, in better agreement with experimental results, except for InP that presents only one peak at the refereed energy position. In this case we argue that possibly the second peak in InP spectra do not appear in experiment due the small number of observed points. 
	
	A common feature observed in all DFT-1/2 spectra depicted in Fig. \ref{fig1} is the decrease of the intensities in relation with both PBEsol and experimental results. As already pointed out the dielectric function is directly related with electronic transitions between valence to conduction bands. Peaks or increase of intensity occurs when the valence and conduction bands exhibit parallel segments in the energy-momentum band dispersion, i.e., the same amount of energy produce excitation of electrons with different momenta in the k-space. In this sense, slight modification of the bands participating of the transitions, in such a way there is a decrease in the parallelism, will cause a reduction of the intensity of absorbance for the corresponding excitation energy. 
	The DFT-1/2 correction in general tends to slightly change the shape of the bands derived from the orbitals for which the self-energy is included. In the most cases only the valence bands are affected, therefore decreasing the parallelism of the bands participating of the transitions which by consequence decrease the intensity of the spectra. Looking the last column of Table \ref{tab2} we can see that except for Ge and GaP the self-energy correction is applied only for orbitals forming the valence band (positive amplitude), resulting in a decrease in intensities. For Ge and GaP we also include self-energy correction for orbitals forming the conduction bands (negative amplitude). For these two cases there is some compensation, i.e., also the bands in the conduction part are modified in such a way that the reduction in intensities of the spectra is less pronounced.  Moreover, one verifies in SM Fig. 1 a similar performance of DFT-1/2+BSE and HSE+BSE approaches. The latter methodology is a cheaper alternative to GW+BSE state-of-art approach, but still orders of magnitude more expansive than former one. 
	
	\section{Conclusions}
	
	In summary, we propose a new procedure to calculate the absorption spectra of semiconductors including quasiparticle energies and excitonic effects with significant reduction in computational effort and in good agreement with experimental results.  It consist in use the DFT-1/2 method to generate an approximated quasiparticle spectra which serves as input to solve the Bethe-Salpeter equation responsible to take into account the excitonic effects. The methodology was tested for a series of semiconductors, namely, Si, Ge, GaAs, GaP, GaSb, InAs, InP and InSb, and result in very good agreement with experimental results. 
	
	The calculated intensities of absorption spectra are slight lower than experimental and standard DFT functional results, which can be explained by the stronger localization of valence bands in the DFT-1/2 picture. Although the intensities of the spectra do not exactly fit the experimental results, their qualitative behavior and the energy positions of the peaks of absorption present a very good agreement. Being DFT-1/2 a low-cost quasiparticle correction method with respect to computational resources, this combination (DFT-1/2 $+$ BSE) can be used to obtain optical spectra including excitonic effects for systems in which GW+BSE approach is unfeasible.

	\begin{acknowledgments}
		We thank the Brazilian funding agency Coordination for Improvement of Higher Level Education (CAPES), PVE Grants No. 88881.068355/2014-01 and No. 88887.145962/2017-00, the National Council for Scientific and Technological Development (CNPq), Grants No. 308742/2016-8 and 306322/2017-0, and the S\~ao  Paulo Research Foundation (FAPESP) for the Grants No. 2006/05858-0 and 2012/50738-3 for the financial supports. We also thank the Scientific Computation National Laboratory (LNCC)  for providing the authors Santos Dumont Supercomputer resources.
		
	\end{acknowledgments}


\begin{thebibliography}{21}
	\expandafter\ifx\csname natexlab\endcsname\relax\def\natexlab#1{#1}\fi
	\expandafter\ifx\csname bibnamefont\endcsname\relax
	\def\bibnamefont#1{#1}\fi
	\expandafter\ifx\csname bibfnamefont\endcsname\relax
	\def\bibfnamefont#1{#1}\fi
	\expandafter\ifx\csname citenamefont\endcsname\relax
	\def\citenamefont#1{#1}\fi
	\expandafter\ifx\csname url\endcsname\relax
	\def\url#1{\texttt{#1}}\fi
	\expandafter\ifx\csname urlprefix\endcsname\relax\def\urlprefix{URL }\fi
	\providecommand{\bibinfo}[2]{#2}
	\providecommand{\eprint}[2][]{\url{#2}}
	
	\bibitem[{\citenamefont{Albrecht et~al.}(1998)\citenamefont{Albrecht, Reining,
			{Del Sole}, and Onida}}]{Albrecht1998}
	\bibinfo{author}{\bibfnamefont{S.}~\bibnamefont{Albrecht}},
	\bibinfo{author}{\bibfnamefont{L.}~\bibnamefont{Reining}},
	\bibinfo{author}{\bibfnamefont{R.}~\bibnamefont{{Del Sole}}},
	\bibnamefont{and} \bibinfo{author}{\bibfnamefont{G.}~\bibnamefont{Onida}},
	\bibinfo{journal}{Phys. Rev. Lett.} \textbf{\bibinfo{volume}{80}},
	\bibinfo{pages}{4510} (\bibinfo{year}{1998}).
	
	\bibitem[{\citenamefont{Benedict et~al.}(1998)\citenamefont{Benedict, Shirley,
			and Bohn}}]{Benedict1998}
	\bibinfo{author}{\bibfnamefont{L.~X.} \bibnamefont{Benedict}},
	\bibinfo{author}{\bibfnamefont{E.~L.} \bibnamefont{Shirley}},
	\bibnamefont{and} \bibinfo{author}{\bibfnamefont{R.~B.} \bibnamefont{Bohn}},
	\bibinfo{journal}{Phys. Rev. Lett.} \textbf{\bibinfo{volume}{80}},
	\bibinfo{pages}{4514} (\bibinfo{year}{1998}).
	
	\bibitem[{\citenamefont{Rohlfing and Louie}(1998)}]{Rohlfing1998}
	\bibinfo{author}{\bibfnamefont{M.}~\bibnamefont{Rohlfing}} \bibnamefont{and}
	\bibinfo{author}{\bibfnamefont{S.}~\bibnamefont{Louie}},
	\bibinfo{journal}{Phys. Rev. Lett.} \textbf{\bibinfo{volume}{81}},
	\bibinfo{pages}{2312} (\bibinfo{year}{1998}).
	
	\bibitem[{\citenamefont{Wirtz et~al.}(2006)\citenamefont{Wirtz, Marini, and
			Rubio}}]{Wirtz2006}
	\bibinfo{author}{\bibfnamefont{L.}~\bibnamefont{Wirtz}},
	\bibinfo{author}{\bibfnamefont{A.}~\bibnamefont{Marini}}, \bibnamefont{and}
	\bibinfo{author}{\bibfnamefont{A.}~\bibnamefont{Rubio}},
	\bibinfo{journal}{Phys. Rev. Lett.} \textbf{\bibinfo{volume}{96}},
	\bibinfo{pages}{126104} (\bibinfo{year}{2006}).
	
	\bibitem[{\citenamefont{Guilhon et~al.}(2019)\citenamefont{Guilhon, Marques,
			Teles, Palummo, Pulci, Botti, and Bechstedt}}]{Guilhon2019}
	\bibinfo{author}{\bibfnamefont{I.}~\bibnamefont{Guilhon}},
	\bibinfo{author}{\bibfnamefont{M.}~\bibnamefont{Marques}},
	\bibinfo{author}{\bibfnamefont{L.~K.} \bibnamefont{Teles}},
	\bibinfo{author}{\bibfnamefont{M.}~\bibnamefont{Palummo}},
	\bibinfo{author}{\bibfnamefont{O.}~\bibnamefont{Pulci}},
	\bibinfo{author}{\bibfnamefont{S.}~\bibnamefont{Botti}}, \bibnamefont{and}
	\bibinfo{author}{\bibfnamefont{F.}~\bibnamefont{Bechstedt}},
	\bibinfo{journal}{Phys. Rev. B} \textbf{\bibinfo{volume}{99}},
	\bibinfo{pages}{161201} (\bibinfo{year}{2019}).
	
	\bibitem[{\citenamefont{Hedin}(1965)}]{Hedin1965}
	\bibinfo{author}{\bibfnamefont{L.}~\bibnamefont{Hedin}},
	\bibinfo{journal}{Phys. Rev.} \textbf{\bibinfo{volume}{139}},
	\bibinfo{pages}{A796} (\bibinfo{year}{1965}).
	
	\bibitem[{\citenamefont{Ferreira et~al.}(2008)\citenamefont{Ferreira, Marques,
			and Teles}}]{Ferreira2008}
	\bibinfo{author}{\bibfnamefont{L.~G.} \bibnamefont{Ferreira}},
	\bibinfo{author}{\bibfnamefont{M.}~\bibnamefont{Marques}}, \bibnamefont{and}
	\bibinfo{author}{\bibfnamefont{L.~K.} \bibnamefont{Teles}},
	\bibinfo{journal}{Phys. Rev. B} \textbf{\bibinfo{volume}{78}},
	\bibinfo{pages}{125116} (\bibinfo{year}{2008}).
	
	\bibitem[{\citenamefont{Ferreira et~al.}(2011)\citenamefont{Ferreira, Marques,
			and Teles}}]{Ferreira2011}
	\bibinfo{author}{\bibfnamefont{L.~G.} \bibnamefont{Ferreira}},
	\bibinfo{author}{\bibfnamefont{M.}~\bibnamefont{Marques}}, \bibnamefont{and}
	\bibinfo{author}{\bibfnamefont{L.~K.} \bibnamefont{Teles}},
	\bibinfo{journal}{AIP Adv.} \textbf{\bibinfo{volume}{1}},
	\bibinfo{pages}{032119} (\bibinfo{year}{2011}).
	
	\bibitem[{\citenamefont{Heyd et~al.}(2003)\citenamefont{Heyd, Scuseria, and
			Ernzerhof}}]{Heyd2003}
	\bibinfo{author}{\bibfnamefont{J.}~\bibnamefont{Heyd}},
	\bibinfo{author}{\bibfnamefont{G.~E.} \bibnamefont{Scuseria}},
	\bibnamefont{and}
	\bibinfo{author}{\bibfnamefont{M.}~\bibnamefont{Ernzerhof}},
	\bibinfo{journal}{J. Chem. Phys.} \textbf{\bibinfo{volume}{118}},
	\bibinfo{pages}{8207} (\bibinfo{year}{2003}).
	
	\bibitem[{\citenamefont{Heyd and Scuseria}(2004)}]{Heyd2004}
	\bibinfo{author}{\bibfnamefont{J.}~\bibnamefont{Heyd}} \bibnamefont{and}
	\bibinfo{author}{\bibfnamefont{G.~E.} \bibnamefont{Scuseria}},
	\bibinfo{journal}{J. Chem. Phys.} \textbf{\bibinfo{volume}{121}},
	\bibinfo{pages}{1187} (\bibinfo{year}{2004}).
	
	\bibitem[{\citenamefont{Heyd et~al.}(2006)\citenamefont{Heyd, Scuseria, and
			Ernzerhof}}]{Heyd2006}
	\bibinfo{author}{\bibfnamefont{J.}~\bibnamefont{Heyd}},
	\bibinfo{author}{\bibfnamefont{G.~E.} \bibnamefont{Scuseria}},
	\bibnamefont{and}
	\bibinfo{author}{\bibfnamefont{M.}~\bibnamefont{Ernzerhof}},
	\bibinfo{journal}{J. Chem. Phys.} \textbf{\bibinfo{volume}{124}},
	\bibinfo{pages}{219906} (\bibinfo{year}{2006}).
	
	\bibitem[{\citenamefont{Perdew et~al.}(2008)\citenamefont{Perdew, Ruzsinszky,
			Csonka, Vydrov, Scuseria, Constantin, Zhou, and Burke}}]{Perdew2008}
	\bibinfo{author}{\bibfnamefont{J.~P.} \bibnamefont{Perdew}},
	\bibinfo{author}{\bibfnamefont{A.}~\bibnamefont{Ruzsinszky}},
	\bibinfo{author}{\bibfnamefont{G.~I.} \bibnamefont{Csonka}},
	\bibinfo{author}{\bibfnamefont{O.~A.} \bibnamefont{Vydrov}},
	\bibinfo{author}{\bibfnamefont{G.~E.} \bibnamefont{Scuseria}},
	\bibinfo{author}{\bibfnamefont{L.~A.} \bibnamefont{Constantin}},
	\bibinfo{author}{\bibfnamefont{X.}~\bibnamefont{Zhou}}, \bibnamefont{and}
	\bibinfo{author}{\bibfnamefont{K.}~\bibnamefont{Burke}},
	\bibinfo{journal}{Phys. Rev. Lett.} \textbf{\bibinfo{volume}{100}},
	\bibinfo{pages}{136406} (\bibinfo{year}{2008}).
	
	\bibitem[{\citenamefont{Perdew et~al.}(2009)\citenamefont{Perdew, Ruzsinszky,
			Csonka, Vydrov, Scuseria, Constantin, Zhou, and Burke}}]{Perdew2009}
	\bibinfo{author}{\bibfnamefont{J.~P.} \bibnamefont{Perdew}},
	\bibinfo{author}{\bibfnamefont{A.}~\bibnamefont{Ruzsinszky}},
	\bibinfo{author}{\bibfnamefont{G.~I.} \bibnamefont{Csonka}},
	\bibinfo{author}{\bibfnamefont{O.~A.} \bibnamefont{Vydrov}},
	\bibinfo{author}{\bibfnamefont{G.~E.} \bibnamefont{Scuseria}},
	\bibinfo{author}{\bibfnamefont{L.~A.} \bibnamefont{Constantin}},
	\bibinfo{author}{\bibfnamefont{X.}~\bibnamefont{Zhou}}, \bibnamefont{and}
	\bibinfo{author}{\bibfnamefont{K.}~\bibnamefont{Burke}},
	\bibinfo{journal}{Phys. Rev. Lett.} \textbf{\bibinfo{volume}{102}},
	\bibinfo{pages}{039902(E)} (\bibinfo{year}{2009}).
	
	\bibitem[{\citenamefont{Kresse and Joubert}(1999)}]{Kresse1999}
	\bibinfo{author}{\bibfnamefont{G.}~\bibnamefont{Kresse}} \bibnamefont{and}
	\bibinfo{author}{\bibfnamefont{D.}~\bibnamefont{Joubert}},
	\bibinfo{journal}{Phys. Rev. B} \textbf{\bibinfo{volume}{59}},
	\bibinfo{pages}{1758} (\bibinfo{year}{1999}).
	
	\bibitem[{\citenamefont{Matusalem et~al.}(2015)\citenamefont{Matusalem,
			Marques, Teles, and Bechstedt}}]{Matusalem2015}
	\bibinfo{author}{\bibfnamefont{F.}~\bibnamefont{Matusalem}},
	\bibinfo{author}{\bibfnamefont{M.}~\bibnamefont{Marques}},
	\bibinfo{author}{\bibfnamefont{L.~K.} \bibnamefont{Teles}}, \bibnamefont{and}
	\bibinfo{author}{\bibfnamefont{F.}~\bibnamefont{Bechstedt}},
	\bibinfo{journal}{Phys. Rev. B} \textbf{\bibinfo{volume}{92}},
	\bibinfo{pages}{045436} (\bibinfo{year}{2015}).
	
	\bibitem[{\citenamefont{Ataide et~al.}(2017)\citenamefont{Ataide, Pel{\'{a}},
			Marques, Teles, Furthm{\"{u}}ller, and Bechstedt}}]{Ataide2017}
	\bibinfo{author}{\bibfnamefont{C.~A.} \bibnamefont{Ataide}},
	\bibinfo{author}{\bibfnamefont{R.~R.} \bibnamefont{Pel{\'{a}}}},
	\bibinfo{author}{\bibfnamefont{M.}~\bibnamefont{Marques}},
	\bibinfo{author}{\bibfnamefont{L.~K.} \bibnamefont{Teles}},
	\bibinfo{author}{\bibfnamefont{J.}~\bibnamefont{Furthm{\"{u}}ller}},
	\bibnamefont{and}
	\bibinfo{author}{\bibfnamefont{F.}~\bibnamefont{Bechstedt}},
	\bibinfo{journal}{Phys Rev B} \textbf{\bibinfo{volume}{95}},
	\bibinfo{pages}{045126} (\bibinfo{year}{2017}).
	
	\bibitem[{\citenamefont{Bechstedt}(2015)}]{Bechstedt2015}
	\bibinfo{author}{\bibfnamefont{F.}~\bibnamefont{Bechstedt}},
	\emph{\bibinfo{title}{{Many-Body Approach to Electronic Excitations}}}, vol.
	\bibinfo{volume}{181} of \emph{\bibinfo{series}{Springer Series in
			Solid-State Sciences}} (\bibinfo{publisher}{Springer},
	\bibinfo{address}{Berlin, Heidelberg}, \bibinfo{year}{2015}).
	
	\bibitem[{\citenamefont{{Mikhail Efimovich Levinshtein}
			et~al.}(1996)\citenamefont{{Mikhail Efimovich Levinshtein}, Rumyantsev, and
			Shur}}]{levinshtein1996handbook}
	\bibinfo{author}{\bibnamefont{{Mikhail Efimovich Levinshtein}}},
	\bibinfo{author}{\bibfnamefont{S.~L.} \bibnamefont{Rumyantsev}},
	\bibnamefont{and} \bibinfo{author}{\bibfnamefont{M.}~\bibnamefont{Shur}},
	\emph{\bibinfo{title}{{Handbook Series on Semiconductor Parameters: Si, Ge, C
				(Diamond), GaAs, GaP, GaSb, InAs, InP, InSb}}} (\bibinfo{publisher}{World
		Scientific}, \bibinfo{year}{1996}).
	
	\bibitem[{\citenamefont{Hahn et~al.}(2005{\natexlab{a}})\citenamefont{Hahn,
			Schmidt, Seino, Preuss, Bechstedt, and Bernholc}}]{Hahn2005a}
	\bibinfo{author}{\bibfnamefont{P.~H.} \bibnamefont{Hahn}},
	\bibinfo{author}{\bibfnamefont{W.~G.} \bibnamefont{Schmidt}},
	\bibinfo{author}{\bibfnamefont{K.}~\bibnamefont{Seino}},
	\bibinfo{author}{\bibfnamefont{M.}~\bibnamefont{Preuss}},
	\bibinfo{author}{\bibfnamefont{F.}~\bibnamefont{Bechstedt}},
	\bibnamefont{and} \bibinfo{author}{\bibfnamefont{J.}~\bibnamefont{Bernholc}},
	\bibinfo{journal}{Phys. Rev. Lett.} \textbf{\bibinfo{volume}{94}},
	\bibinfo{pages}{1} (\bibinfo{year}{2005}{\natexlab{a}}), ISSN
	\bibinfo{issn}{00319007}.
	
	\bibitem[{\citenamefont{Hahn et~al.}(2005{\natexlab{b}})\citenamefont{Hahn,
			Seino, Schmidt, Furthm{\"{u}}ller, and Bechstedt}}]{Hahn2005}
	\bibinfo{author}{\bibfnamefont{P.~H.} \bibnamefont{Hahn}},
	\bibinfo{author}{\bibfnamefont{K.}~\bibnamefont{Seino}},
	\bibinfo{author}{\bibfnamefont{W.~G.} \bibnamefont{Schmidt}},
	\bibinfo{author}{\bibfnamefont{J.}~\bibnamefont{Furthm{\"{u}}ller}},
	\bibnamefont{and}
	\bibinfo{author}{\bibfnamefont{F.}~\bibnamefont{Bechstedt}},
	\bibinfo{journal}{Phys. Status Solidi Basic Res.}
	\textbf{\bibinfo{volume}{242}}, \bibinfo{pages}{2720}
	(\bibinfo{year}{2005}{\natexlab{b}}), ISSN \bibinfo{issn}{03701972}.
	
	\bibitem[{\citenamefont{Aspnes and Studna}(1983)}]{Aspnes1983}
	\bibinfo{author}{\bibfnamefont{D.~E.} \bibnamefont{Aspnes}} \bibnamefont{and}
	\bibinfo{author}{\bibfnamefont{A.~A.} \bibnamefont{Studna}},
	\bibinfo{journal}{Phys. Rev. B} \textbf{\bibinfo{volume}{27}},
	\bibinfo{pages}{985} (\bibinfo{year}{1983}).
	
\end{thebibliography}
\end{document}